\documentclass[12pt,preprint]{aastex}
\slugcomment{draft version}
\RequirePackage{longtable}%

\shorttitle{Cross-correlation and time lag of 4U 1735-44}
\shortauthors{Y.J. Lei et al.}

\begin{document}

\title{Evolution of cross-correlation and time lag of 4U 1735-44 along the branches}

\author{Ya-Juan Lei\altaffilmark{1}, Hao-Tong Zhang\altaffilmark{1}, Cheng-Min Zhang\altaffilmark{1},Jin-Lu Qu\altaffilmark{2}, Hai-Long Yuan\altaffilmark{1}, Yi-Qiao Dong\altaffilmark{1}, Yong-Heng Zhao\altaffilmark{1},De-Hua Wang\altaffilmark{3}, Hong-Xing Yin\altaffilmark{4}, Li-Ming Song\altaffilmark{2}}

\altaffiltext{1}{Key Laboratory of Optical Astronomy, National Astronomical Observatories, Chinese
Academy of Sciences, Beijing 100012, P.R. China; leiyjcwmy@163.com}

\altaffiltext{2}{Particle Astrophysics Center, Institute of High
Energy Physics, Chinese Academy of Sciences, Beijing 100049, P.R.
China}

\altaffiltext{3}{Astronomy Department, Beijing Normal University, Beijing 100875, P.R.
China}

\altaffiltext{4}{School of Space Science and Physics, Shandong University, Weihai, 264209, P.R.
China}

\begin{abstract}

We analyze the cross-correlation function between the soft and  hard
X-rays of  atoll source 4U 1735-44 with {\it RXTE} data, and find
the anti-correlated soft and hard time lags  of about hecto-second.
On the  island state, the observations do not show any  obvious
correlations, and most  observations of banana branch show 
positive correlation.  However, the anti-correlations  are  detected
on the upper banana branch. These results   are different from those
of Z sources (Cyg X-2, GX 5-1), where the  anti-correlation is
detected in the low luminosity states,  then the lag  timescales  of
 both  this atoll and  Z sources are found to be similar, at the magnitude of
 several tens to hundreds of seconds.  As a comparison, it is noted that  the
 anti-correlated lags of thousand-second  have been  reported from the several black hole
candidates in their intermediate states.
    Finally,  we compare the correspondent  results of  atoll source  4U 1735-44
     with those observed in black hole candidates and Z sources,
and the possible origins of the anti-correlated time lags are
discussed.

\end{abstract}


\section{Introduction}

 The timing and spectral analysis of the X-ray emission of
the compact objects, neutron stars (NSs) or black holes (BHs),  can
exhibit  the accreting flow properties of low-mass X-ray binaries
(LMXBs) \citep[see, e.g.,][]{van06,liu07a}.
 Based on the  X-ray spectral properties, which are shown in
 the defined color-color diagram (CCD) or hardness-intensity diagram (HID),
 the NS  LMXBs can be classified into two subtypes, Z
sources with high luminosity and atoll sources with low luminosity \citep{has89}.
  The main difference between atoll and Z
sources might be the accretion rate and  magnetic field, where Z
  source shares the higher  accretion rate and slightly
stronger magnetic field than those  of atoll source \citep{zha07}.
The atoll sources cover a wider range of luminosity,
whereas  Z sources  usually  appear as the Eddington limited
luminosities \citep[e.g.,][]{for00}.
Z (atoll) source  traces  out  a Z-shape track (island + banana
pattern) within a few days (days to weeks).
In CCD,  the Z-track  consists of three main branches, horizontal
branch (HB), normal branch (NB) and flaring branch (FB) \citep{has89,
has90}. The source evolves continuously along the Z-track, with each
position related to the different mass accretion rate. The accretion
rate increases from HB to NB and reaches a maximum at the right end
of the FB \citep{hasetal90}.
However, recently, the results of studying the source XTE J1701-462
suggest that the motion along the Z branches  might be corresponding
to the roughly constant accretion rate \citep{lin09, hom10}.
As  for the CCD pattern of atoll source,  it follows the similar
correspondences  to those of Z source, from island state (IS)   to
the lower left  banana (LLB), lower banana (LB)  and upper banana
(UB) \citep{van00, alt08}.
 In the atoll source, the count rates increase with the inferred
accretion rate $\dot{M}$, which  increases from the IS  to the UB.  Some atoll
sources show the Quasi-periodic oscillations (QPOs)  in the  LLB
(\citealt{van00, van06, bel07}). QPOs are usually not detected in
the IS, which may be on account of the  low sensitivity at the low
count rates.
%but in one island in 4U 0614$+$09 the undetected lower kHz
%peak is really much weaker than at higher inferred $\dot{M}$
%\citep{van95, men97}.

The spectral characteristics of Black hole X-ray binaries (BHXBs)
usually exhibit five distinct  states, quiescent state, low/hard
state(LHS), intermediate state (IMS), high/soft state (HSS), and
very high state (VHS) \citep{rem05, rem06, bel10}. The IMS is
complex, whose luminosity could not be determined solely by the mass
accretion rate \citep{yu07, yu09}.  Although the luminosity of VHS
is much higher than that of IMS, both states share the  similar
spectral and timing behaviors, which
 can be taken as the same state that  represents  the transitions
between LHS and HSS \citep{don07}.

%For each type of source, several spectral/timing states are
%identified to be associated with the inner accretion flows.

The timing properties of X-ray binary are related to the spectral
states i.e., the positions  of CCD.  The cross-correlation function
of the soft and hard X-ray photons  can be used for analyzing the
relation of different energy bands, which is helpful for
understanding the radiative and geometrical structure of the
accretion disk for different spectral states.  The cross-correlated
soft and hard X-ray lags  have been investigated for  Z sources and
BHXBs.
For some BHXBs, the anti-correlation of a few hundred seconds are
often seen in the steep power-law (SPL) or intermediate states \citep{cho05, sri07,
sri09, sri10}, and these timescales of the lags could be ascribed to
the viscous timescales of matter flows in the optically thick
accretion disks \citep{cho04}. Such anti-correlated lags ($\sim$
several 10-100 s) are also detected on the HB and upper NB of two Z
sources  (Cyg X-2, GX 5-1) \citep{lei08, sri12}. The simulation by
\citet{sri10} indicates that
  the phenomena of observed anti-correlations should not arise from
  the   spurious nature, but the reality of physical processes.

In some aspects, the atoll source and BH system
share the  similar spectral characteristics  \citep{dis06}, thus we
select an atoll source, 4U 1735-44, to study  its  evolution of
cross-correlation and time lag between the soft and hard X-rays.
This source  is classified as an atoll source with a relatively high
luminosity \citep{sma86, seo97}, the distance of which  is measured
to be about 8.5 kpc
 \citep{gal08}. The binary parameters of 4U 1735-44 have been observed with
  the orbital period of 4.654 hr \citep{cor86} and   its companion  mass
  of  0.53$\pm$0.44 $M_{\odot}$ \citep{cas06}. Its  X-ray spectra
   are analyzed by \citet{ng10} and \citet{mue11}, with  an iron
emission line detected. In addition, the kHz QPOs are discovered at
the relatively lower mass accretion rates \citep{for98, wij98}.
In detail,  we investigate whether the anti-correlation between the
soft (2-3.3 keV)  and hard (12-30 keV) X-rays exists in  4U 1735-44.
As a further step, we compare the results of atoll with those of Z
sources and BHXBs. The organization of the paper is described below.  Section 2 is about
the observations and data analysis of {\it RXTE},  and the results
on the cross-correlation between the soft and hard X-rays  are
analyzed in section 3. The  discussions and  summary are written in section 4.

\section{Observation and Data Reduction}

We analyze all pointed observations of  the Proportional
Counter Array (PCA) on board the {\it RXTE} satellite. The PCA
consists of 5 non-imaging, coaligned Xe multiwire proportional
counter units (PCUs). In the work, only PCU2 data are adopted,
which is the best calibrated unit and has the longest observational duration.
After excluding  the data  segments less  than  2000 s,    we
extract   the light curves   from the data of standard 2
 mode   with bin size of 16 s.
With the XRONOS tool ``{\it crosscor}'', we  estimate the coefficient of the cross-correlation
between the soft (2-3.3 keV) and the hard (12-30 keV) X-rays.
%
%The ``{\it crosscor}'' can be used to compute  the coefficient of
% cross-correlation.

 Similar to the previous publication  about the  cross-correlation
   for Z source Cyg X-2,
 we divide the cross-correlation results  into three groups,
 the positive, ambiguous and anti-correlated \citep[also see][]{lei08}.
 We can  obtain the time lag by  fitting  the  anti-correlated part  with an
inverted Gaussian function.
After  subtracting the  background, the  light curves and
anti-correlations between soft and hard X-rays, accompanying with
the pivoting spectra from the data with different hardness ratios,
are shown in Figure 1, where we define the hard and soft regions
according to the hardness ratios.

In order to study the source evolution on the CCD of 4U 1735-44,  we
define the soft and the hard colors  as the count-rate ratios 3.5-6.0
keV/2.0-3.5 keV and 9.7-16 keV/6.0-9.7 keV \citep[also
see][]{obr04}. The light curves are extracted from the 2.0-3.5 keV,
3.5-6.0 keV, 6.0-9.5 keV and 9.5-16 keV, respectively, with the
background subtracted.
By the corresponding Crab values, we correct the affections of the
different gain epochs and the gain change \citep{van05}.
Figure 2 shows the revised CCDs with the bin size of 512 s. For
spectral analysis, the PCA background subtraction is carried out
using the latest versions of the appropriate background models, and
a 0.6\% systematic error is added to the spectra to account for the
calibration uncertainties. As usual, the spectral fitting software
$XSPEC$ is used.

\section{Results}

\subsection{The results of cross-correlations  and  their  distribution on CCD}

With the data  of {\it RXTE},  for the atoll source 4U 1735-44, we
analyze the cross-correlation between the soft (2-3.3 keV) and  hard
(12-30 keV) X-rays.
For each segment of the light curves, we analyze the
cross-correlation.  
The anti-correlations have been detected in  the light curves
of 12 observations. We apply  the letter `a'  to mark the segment of
the anti-correlation.
The ranges of hard and soft  time lags for the anti-correlation are
from a few tens to hundred seconds.
There are four regions in Figure 2 that shows the   CCD of all
analyzed  observational data.
The data in region `I', `II', `III' and  `IV'  are mainly related to
the IS,  LLB,   LB and  UB, respectively.

In Table 1, we list the observations with
anti-correlation detected, the cross-correlation coefficients and delay
times.  From  Table 1 and Figure 2, we notice that most of the
anti-correlated observations are located at the UB, where  show the
7 observations with anti-correlated hard X-ray lags, 3 observations
with anti-correlated soft X-ray lags and 2 observations with
unobvious time lag.
The delay times are similar to those found in the Z sources Cyg X-2
and GX 5-1 \citep{lei08, sri12}.     There is no relation detected
between the delay times and the location in region `IV' of Figure 2.

%Figure 2 shows  the CCD of all analyzed observations of 4U 1735-44.

From Figure 2 and Table 2, we can see that, in the region `I',
ambiguous correlated data are dominated,  then the great majority of
positively correlated data dominate  in the regions `II' and `III',
and in the region `IV', all the three correlations are found  in the
equal weight.
Averagely, the correlation coefficients  of  the region `II'  are
mostly less than those of region `III'.  Obviously, most
anti-correlated data occur  in regions `IV'.
In addition,  for the observations with the anti-correlations,
the spectral evolution of  the soft and hard regions of
the light curves is shown in Figure 1. However, the energy ranges of the pivoting are not obviously
evolutional along the locations of the CCD.
There is not yet found the  pivoting in the spectral evolution of
positive-correlation between the soft and hard X-rays,  which is
consistent with the results of \citet{lei08} and \citet{sri12}.

%The hard and the soft X-rays are 0\% anti-correlated and
%19\% positively correlated in region `I'; 0\%  anti-correlated and
%90\% positively correlated in region `II'; $\sim$2\%  anti-correlated and
%$\sim$94\% positively correlated in the regions `III',  30\%  anti-correlated and
%28\% positively correlated in region `IV' (Tab. 2).
%In the region `I', ambiguous data are the main.
%Obviously,  most anti-correlated data are  in regions `IV',
%and the great majority of positively correlated data  are in the regions `II' and `III'.

\subsection{Spectral variation }
It is known from  the previous work  that  the  position  in the CCD
generally is well correlated to its spectral/timing state
\citep[see, e.g.,][]{van06}.
For studying the spectral evolution during the positive and
anti-correlation between soft and hard X-rays, we analyze the
spectra of the segments a, b, c and d (account rates decrease from a
to d) of ObsID 91025-01-04-03, separately.
For segment a, the cross-correlation between soft and hard X-rays
shows anti-correlation,  and for segment b  the cross-correlation
shows strong positive correlation, and,  for segment c and d, the
cross-correlations show obviously positive correlation but with a
lower correlation coefficients than that of segment b.

These spectra can be fitted by the multi-color disk model ($diskbb$
in XSPEC) plus black body ($bbody$ in XSPEC) \citep[e.g.,][]{wen11}.
A gaussian line with the center energy fixed at
 6.6 keV and the width fixed at 0.5 keV
is added to describe the iron line.
The photo-electric absorption is modelled with $wabs$ in XSPEC. Due
to the absence of low energy observation of PCA, the absorption
column is fixed at the value of $N_{\rm H}$ = 3.4 $\times$ $10^{21}$
cm$^{-2}$ \citep{chr97}.

Figure 3 and Table 3 show the fitting results, from segment a to d,
as the X-ray flux decreasing, both the disk temperature and black
body temperature decrease independently.
We can see that, with the increasing of the flux (from d to a),  the
inner disk radius($\varpropto$ the normalization)  decreases,  as
the region of black body emission ($\varpropto$ the normalization)
increases.
In addition, for segments a and d, both the soft and hard components
dominate with the similar weights,  the spectral property of  which
is corresponding to the SPL states of BHXBs \citep[also
see][]{wen11}.
For segments b and c, the soft components dominate
the spectral characteristic, which is similar to the HSS of BHXBs
\citep[also see][]{sri10, wen11}.

\section{Summary and Discussion}

We  have detected the anti-correlation between soft (2-3.3 keV) and
hard (12-30 keV) X-rays from the atoll source 4U 1735-44 in 12 {\it RXTE} observations.
The results show that, the observations with anti-correlation are located at
the UB,  and  the observations with positive correlation  mostly
appear in the LLB and LB.  Most observations in the IS show the
ambiguous correlation.
The pivoting of the wide-band X-ray spectrum
is also detected in all the observations with  the anti-correlation detected.
The energies of pivoting are generally  lower than those detected in
Z sources, and are  not related to their  location of CCD (see Table 1).

The results from the atoll source 4U 1735-44 are  different
from those of Z sources \citep{lei08, sri12}.
 Most of the
anti-correlated observations of Z sources locate at HB and upper NB,
 which correspond to ``the low-hard state'',  however  the
anti-correlated observations detected in the atoll source 4U 1735-44 locate at UB
which corresponds to high luminosity state.
The anti-correlated time lag scales of  Z and atoll sources are
similar, and the observations with anti-correlated hard X-ray lags
are more than those of soft X-ray lags for both Z  and atoll
sources.
For Z source, if the accretion rate is the primary parameter to
determine the location on CCD, the anti-correlation could be
corresponding to the lower mass accretion.  However, the mass
accretion rate may be constant for XTE J1701-462 during the Z-track
\citep{lin09}. Similar to Z sources, there is no obvious relation
between the observed delays and corresponding positions on CCD for
the atoll source 4U 1735-44.

The anti-correlations have been  detected in a few BHXBs with a few
hundred to thousand seconds lags with the cross-correlation function
of soft and hard X-rays, which  suggests that the accretion disk
could be truncated \citep{cho05, sri07, sri09, sri10}.
For two Z sources, Cyg X-2 and GX 5-1, the anti-correlation between
soft and hard X-rays favors the truncated accretion disk geometry
\citep{lei08, sri12}.
The truncated accretion disk model is often used to explain the
spectral and temporal features of the hard states observed in BHXBs
\citep{don07}, however some observational results show that the disk
may not be truncated \citep{mil06, ryk07}.
The iron line is corresponding to the
innermost disk regions \citep{ste90}.
 \citet{don06} suggest
that truncated disk models are consistent with  the detected smeared  iron
emission line. \citet{don10} show  that the iron line is instead narrow and
   consistent with the truncated disk geometry in GX 339-4.
 For the atoll source 4U 1705-44, the similar profile of iron emission
 line is found in the high/soft states, which suggests that the accretion
 disk could be also truncated
  in the high/soft states of atoll source \citep{dis09}.
  The anti-correlation detected in the high luminosity state should
  also   imply   that the truncated disk   occurs there.

For BHXBs, \citet{esi97} propose that an accretion flow around a BH
consists of two zones with a transition radius,  an inner
advection-dominated accretion flow (ADAF) and an outer standard thin
disk.
In generally, in the LH state  the disk is assumed to be truncated
at a large radial distance,  and  in the  HS  state it becomes
non-truncated.
Another suggestion is that the accretion disk is truncated very
close to the black hole in the IM state/SPL state from the spectral
and temporal results of the study of various black hole sources
\citep[e.g,][]{sri10}.
The anti-correlation detected in the BHXBs  implies that
the  ADAF condenses and expanses to the inner disk,  and therefore the
hard X-ray flux increases and  soft  X-ray flux decreases. The
IM state is probably  the  spectral state  where the condensation
of the inner hot matter can transform into an inner disk
\citep{liu07b, mey07, sri10}.

The NS has a solid surface and magnetic field,  which is different
from those of BH. However,  if the luminosity of NS LMXB  increases
above a critical luminosity, its  magnetosphere may go inside ISCO,
which may account for the same disk evolutionary pattern for BHXBs  \citet{wen11}.
Our results show that the anti-correlation is detected in the UB which is
corresponding to the max flux for the atoll source 4U 1735-44.
According to the fitting results of spectra and X-ray luminosity,
 the LB could be corresponding to the HSS of black hole,
 and the UB is in accordance with the  VHS.
Although the  different luminosity  for IMS and VHS of BHXBs, both
states share the similar spectral and timing behavior, and they are
often thought  as the same state and stand for the transitions
between LHS and HSS \citep{don07}.
The anti-correlations occur  at SPL of BHXBs and at UB of NS system (atoll),
so we can  take  the UB of atoll  to be similar to SPL of BH system.

Furthermore, we notice that the soft and hard X-rays are not
correlated in the IS state for atoll source, which could be due to the fact
that the  accretion disk boundary  is far away from   the hot corona
where emits  the high energy photons.
With the accretion rate increasing, the accretion disk moves inward
with more soft seed photons produced,  at the same time, the ADAF becomes be condensed  and produces
more hard photons, therefore the positive-correlation occurs, implying
the increases of  both the soft and hard X-rays.
When the luminosity increases above a critical luminosity, the ADAF
extends outside to the form of the  corona,  which expands quickly, so
the soft X-rays decrease with the hard X-rays increasing.
In the UB,   the accretion disk extends down to ISCO and could be
truncated there, and the anti-correlations can be detected.
 Our spectral analysis suggests that with the
increasing of the accretion rate, the inner radius of the accretion
disk decreases, which is consistent with BHXBs that the accretion
disk could be truncated very close to the compact object
\citep{sri10}.

The anti-correlated lags are detected and range from a few tens to
hundred seconds in atoll source 4U 1735-44, which  are similar to
those found in Z sources \citep{lei08, sri12}, but  time lags are
lower than those in BHXBs \citep[e.g.,][]{cho04, sri09}.
Generally,  the hard X-ray emission could come from the Comptonizing
of the soft seed photons.
For BHXBs, the soft seed photons could only come from the accretion disk.
For NS LMXB  systems,    the soft seed photons could be produced
from the surface of NS and/or the accretion disk.
The  Comptonizing might occur at a hot corona, a hot flared inner
disk, or even the boundary layer between the NS and accretion disk
\citep[also see][]{pop01, qu01}.
The time scale of only Comptonization process is expected to be about
$\lesssim$ 1 second,  therefore the observed several hundred-second
lags  could be the other  processes\citep{has87, now99,
bot99}.   We infer that  this  anti-correlated  lag time scale of
about several hundred seconds would be related to the change of the disk
structure. The different anti-correlated lag timescale  implies that the disk
inner radius occurs at various locations, which can also affect
corona region. \citet{sri12} suggest that the observed lags could be due to the
change in the size of the Comptonizing region.
Some authors  suggest that the lag time of several hundred seconds
is comparable to the viscous timescale  during which the various
physical and radiative processes change \citep[e.g.,][]{cho04,
lei08, sri12}.
However, above  model cannot explain the anti-correlated soft X-ray
lag, which can  be explained   that a fluctuation produced from the
innermost accretion disk could propagate to the outer accretion disk
and affect the soft X-ray emission regions.
The  fluctuation  of  accretion disk   by the  thermally unstable,
which moves outward from the inner to outer disk,  perturbs  the
outer material which can produce  the soft X-ray lag at a timescale of a
few hundred seconds \citet{li07}.

\acknowledgements 
I would like to thank the referee. 
This research has made use of data obtained
through the high-energy Astrophysics Science Archive Research Center
Online Service, provided by the NASA/Goddard Space Flight Center. We
acknowledge the RXTE data teams at NASA/GSFC for their help. This
work is subsidized by the Special Funds for Major State Basic
Research Projects and by the  Natural Science Foundation of China
for supports with numbers 2012CB821800 and  2009CB824800 and NSFC
numbers 10903005, 11173034, 11173024 and 10903007.

\clearpage

\begin{figure*}
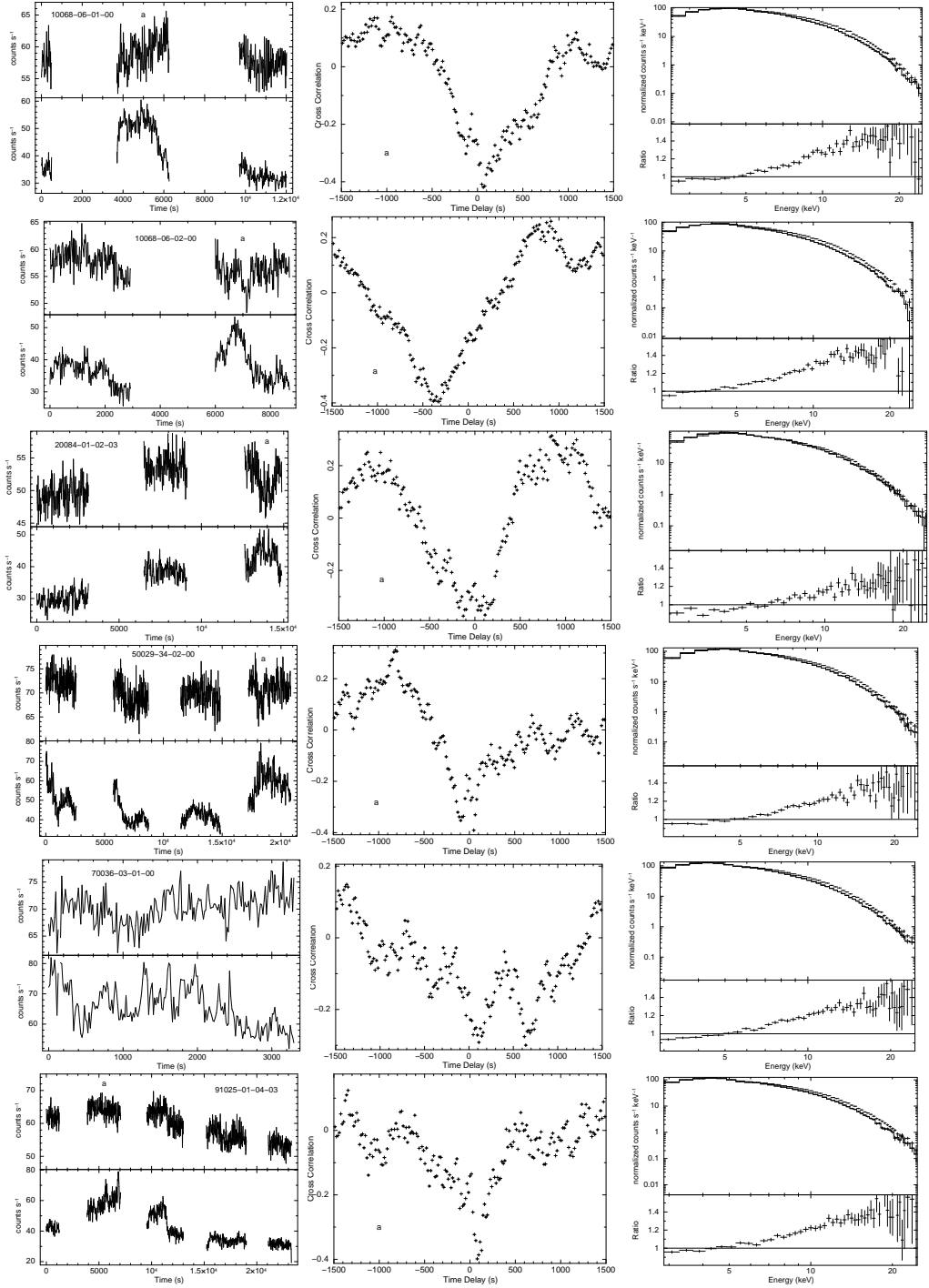

\begin{center}
\includegraphics[width=3.1cm,angle=270,clip]{f1a.ps}
\includegraphics[width=3.1cm,angle=270,clip]{f1b.ps}
\includegraphics[width=3.1cm,angle=270,clip]{f1c.ps}
\includegraphics[width=3.1cm,angle=270,clip]{f1d.ps}
\includegraphics[width=3.1cm,angle=270,clip]{f1e.ps}
\includegraphics[width=3.1cm,angle=270,clip]{f1f.ps}
\includegraphics[width=3.1cm,angle=270,clip]{f1g.ps}
\includegraphics[width=3.1cm,angle=270,clip]{f1h.ps}
\includegraphics[width=3.1cm,angle=270,clip]{f1i.ps}
\includegraphics[width=3.1cm,angle=270,clip]{f1j.ps}
\includegraphics[width=3.1cm,angle=270,clip]{f1k.ps}
\includegraphics[width=3.1cm,angle=270,clip]{f1l.ps}
\includegraphics[width=3.1cm,angle=270,clip]{f1m.ps}
\includegraphics[width=3.1cm,angle=270,clip]{f1n.ps}
\includegraphics[width=3.1cm,angle=270,clip]{f1o.ps}
\includegraphics[width=3.1cm,angle=270,clip]{f1p.ps}
\includegraphics[width=3.1cm,angle=270,clip]{f1q.ps}
\includegraphics[width=3.1cm,angle=270,clip]{f1r.ps}
\caption{The lightcurves (left), cross-correlations between the soft (2-3.3
keV) and hard (12-30 keV) X-rays (middle), the X-ray spectra (right) of the ObsID for
hard and soft regions of the light curves are shown. } \label{fig1}
\end{center}
\end{figure*}
\clearpage
\pagestyle{empty}
\begin{center}
\vspace*{-27mm}
\includegraphics[width=3.1cm,angle=270,clip]{f1s.ps}
\includegraphics[width=3.1cm,angle=270,clip]{f1t.ps}
\includegraphics[width=3.1cm,angle=270,clip]{f1u.ps}
\includegraphics[width=3.1cm,angle=270,clip]{f1v.ps}
\includegraphics[width=3.1cm,angle=270,clip]{f1w.ps}
\includegraphics[width=3.1cm,angle=270,clip]{f1x.ps}
\includegraphics[width=3.1cm,angle=270,clip]{f1y.ps}
\includegraphics[width=3.1cm,angle=270,clip]{f1z.ps}
\includegraphics[width=3.1cm,angle=270,clip]{f1aa.ps}
\includegraphics[width=3.1cm,angle=270,clip]{f1ab.ps}
\includegraphics[width=3.1cm,angle=270,clip]{f1ac.ps}
\includegraphics[width=3.1cm,angle=270,clip]{f1ad.ps}
\includegraphics[width=3.1cm,angle=270,clip]{f1ae.ps}
\includegraphics[width=3.1cm,angle=270,clip]{f1af.ps}
\includegraphics[width=3.1cm,angle=270,clip]{f1ag.ps}
\includegraphics[width=3.1cm,angle=270,clip]{f1ah.ps}
\includegraphics[width=3.1cm,angle=270,clip]{f1ai.ps}
\includegraphics[width=3.1cm,angle=270,clip]{f1aj.ps}\\[5mm]
\centerline{Fig. 1. --- Continued.}
\end{center}
\clearpage

\begin{figure*}
\begin{center}
\includegraphics[width=14cm,angle=0,clip]{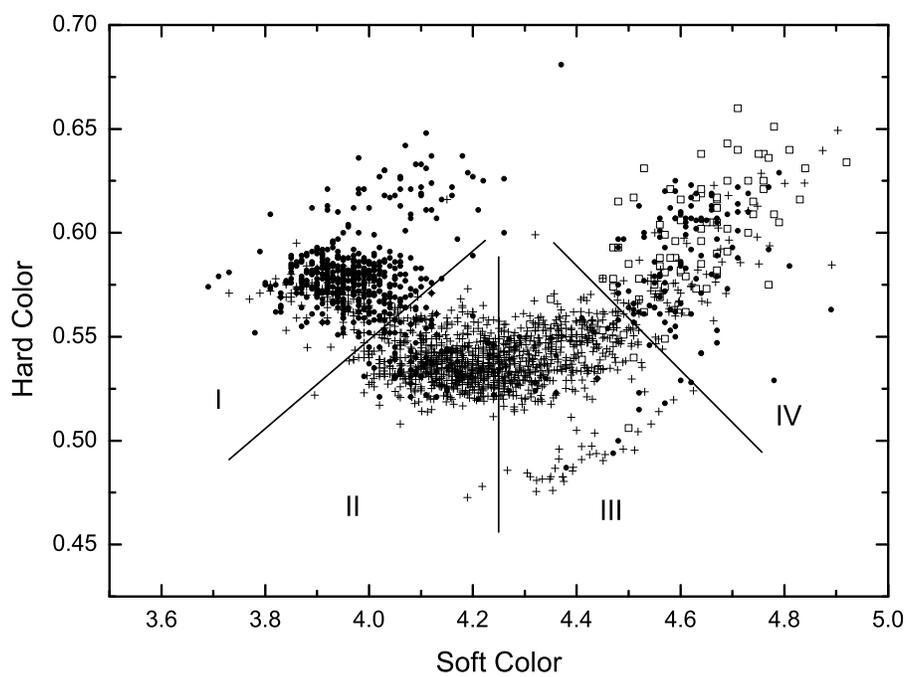}

\caption{CCD of all the  analyzed  observations is divided into four regions,
where squares (dots, crosses) stand for the data with anti-correlation (ambiguous, positive correlation), respectively.
}
\label{fig2}
\end{center}
\end{figure*}

\begin{table*}
\footnotesize
%%\tiny
\caption{\bf ~~Parameters of ObsIDs where the anti-correlations  are detected}
\scriptsize{} \label{table:1}
\newcommand{\m}{\hphantom{$-$}}
\newcommand{\cc}[1]{\multicolumn{1}{c}{#1}}
\renewcommand{\tabcolsep}{0.6pc} % enlarge column spacing
\renewcommand{\arraystretch}{1.2} % enlarge line spacing
\medskip
\begin{center}
\begin{tabular}{c c c c c c c c c}
\hline
     & Date & & & Delay(Error)& Hardness Ratio&Energy of pivoting \\
%\cline{5-6}
     ObsID      & Year-Month-Day        & Location   & CC    & (s)
      &  & (keV) \\
\hline

  10068-06-01-00 & 1996-09-01   & III/IV &  -0.34$\pm$0.02          & 161$\pm$20        &  4.69/0.60    &  $\sim$4.5 \\
  10068-06-02-00 & 1996-09-04   & III/IV &  -0.37$\pm$0.01          &-350$\pm$10        &  4.562/0.568  & $\sim$3.7 \\
  20084-01-02-03 & 1997-09-01   & IV &  -0.34$\pm$0.02         &-25.9$\pm$24.9      &  4.636/0.592  & $\sim$6\\
  50029-34-02-00 & 2000-08-14   & IV &  -0.29$\pm$0.03          &-16.8$\pm$21.0     &  4.73/0.608  & $\sim$5\\
  70036-03-01-00 & 2002-06-11   & IV &  -0.20$\pm$0.02         &317$\pm$8           &  4.827/0.616  & $\sim$5.1 \\
  91025-01-04-03 & 2007-01-20   & IV &  -0.30$\pm$0.03         &88.7$\pm$11         &  4.665/0.599  & $\sim$4.3\\
  91025-01-04-08 & 2007-01-23   & III/IV &  -0.27$\pm$0.03         & 63.4$\pm$33.4      &  4.64/0.609  & $\sim$4 \\
  91025-01-06-01 & 2007-06-22   & III/IV &  -0.27$\pm$0.04         &-113$\pm$18         &  4.571/0.589  & $\sim$4.2\\
  91025-01-06-02 & 2007-06-23   & IV &  -0.27$\pm$0.03        &67.6$\pm$23.2        &  4.61/0.607  & $\sim$5.1\\
  91025-01-06-08 & 2007-06-27   & III/IV &  -0.25$\pm$0.02        &152$\pm$28           &  4.584/0.601  & $\sim$3.9\\
  91025-01-06-09 & 2007-06-28   & III/IV &  -0.30$\pm$0.02       &154$\pm$24            &  4.495/0.593  & $\sim$4.7\\
  91025-01-09-05 & 2008-01-13   & IV &  -0.18$\pm$0.02        &-192$\pm$49          &  4.574/0.596  & $\sim$7 \\

\hline
\end{tabular}
\end{center}
\footnotesize
\end{table*}

\begin{table*}
\label{table:2}
\begin{center}
\caption{\bf ~~Percentage of positive and anti-correlation for each
region of Fig. 1}
\medskip
\begin{tabular}{l c c c c}
\hline
       & I      & II & III &  IV \\
\hline
  positive-correlation &  19\% & 90\% & 94\% &28\%\\
\hline
    anti-correlation   &  0\%  & 0\%  & 2\% &30\%\\

\hline
   ambiguous           & 81\%  &  10\% &4\% & 42\%\\

\hline
\end{tabular}
\end{center}
\end{table*}

\clearpage

\begin{figure*}
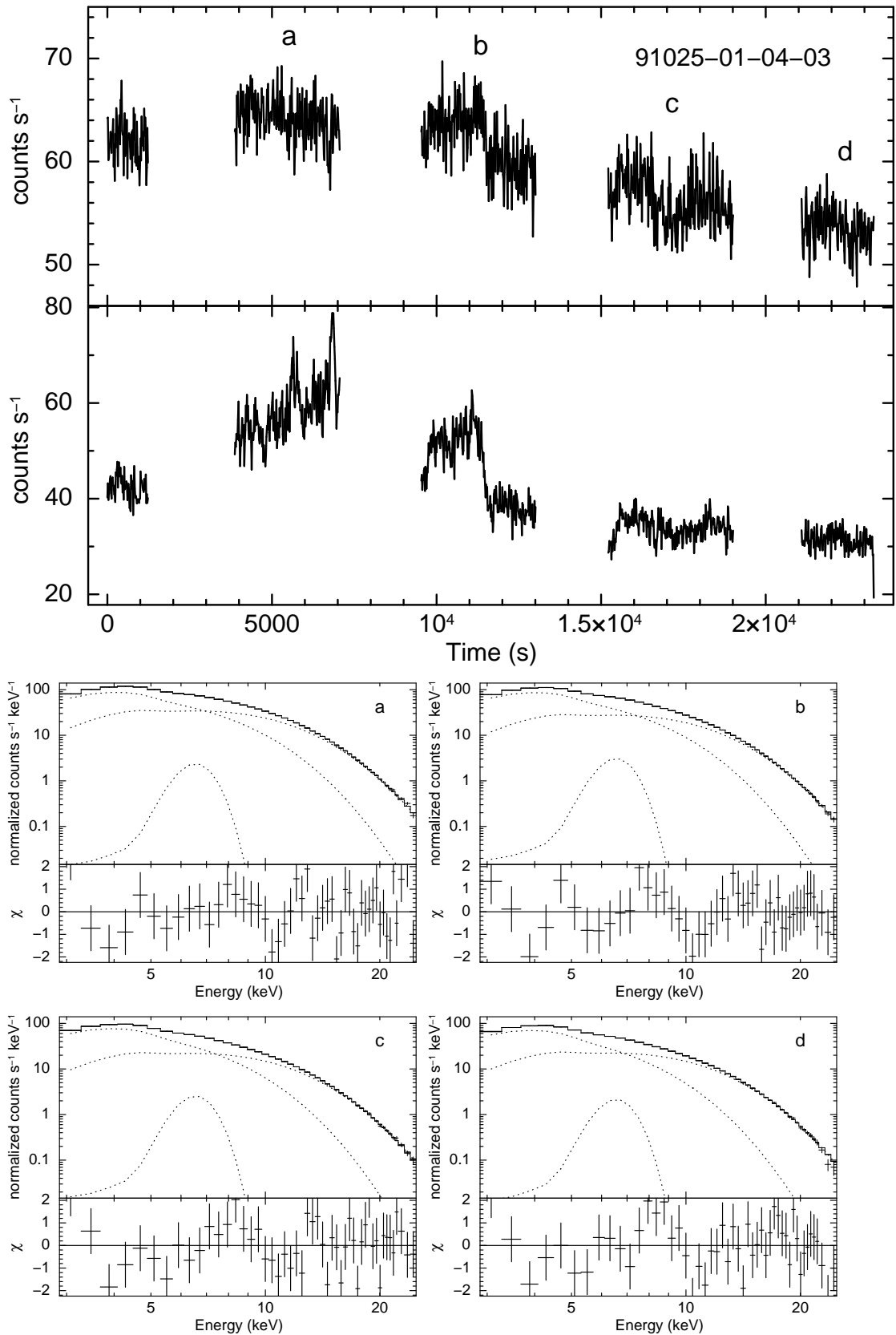

\includegraphics[width=11cm,angle=270,clip]{f3a.ps}
\includegraphics[width=5.5cm,angle=270,clip]{f3b.ps}
\includegraphics[width=5.5cm,angle=270,clip]{f3c.ps}\\
\includegraphics[width=5.5cm,angle=270,clip]{f3d.ps}
\includegraphics[width=5.5cm,angle=270,clip]{f3e.ps}
\caption{Top: The light curves of soft and hard X-rays of ObsID 91025-01-04-03.
Middle and Bottom: The unfolded spectrum for each segment is plotted along with its model components.}
\label{fig3}
\end{figure*}

\begin{table*}
\footnotesize
%%\tiny
\caption{\bf ~~The fitting parameters of the spectra of ObsID 91025-01-04-03}
\scriptsize{} \label{table:3}
\newcommand{\m}{\hphantom{$-$}}
\newcommand{\cc}[1]{\multicolumn{1}{c}{#1}}
\medskip
\begin{center}
\begin{tabular}{c c c c c c c c }
\hline
    Parameters & $kT_{in}$(keV)& $N_{disk}$ & $kT_{bb}$ & $N_{bb}$ & $Flux_{diskbb}/Flux_{total}$ &$Flux_{bb}/Flux_{total}$ &$\chi^2$(d.o.f)\\
\cline{5-6}

\hline
 a            &  1.95$\pm$0.09        &     18.6$\pm$2.8      &       2.72$\pm$0.05    &     0.051$\pm$0.003 &  4.07/8.23 (49.5\%) & 4.12/8.23 (50.1\%) &  1.20(45)  \\
 b            &  1.86$\pm$0.07        &     21.4$\pm$2.6      &       2.68$\pm$0.05    &     0.040$\pm$0.002 &  3.91/7.22 (54.2\%) & 3.26/7.22 (45.2\%) & 0.84(45) \\
 c            &  1.73$\pm$0.05        &     25.7$\pm$2.7      &       2.65$\pm$0.04    &     0.031$\pm$0.002 &  3.32/5.91 (56.2\%) & 2.55/5.91 (43.1\%) & 1.20(45)  \\
 d            &  1.63$\pm$0.05        &     39.9$\pm$3.6      &       2.59$\pm$0.03    &     0.031$\pm$0.002 &  2.75/5.51 (49.9\%) & 2.66/5.51 (48.3\%)& 1.31(45) \\

\hline
\end{tabular}
\end{center}
\footnotesize {The flux in unit of 10$^{-9}$ ergs cm$^{-2}$ s$^{-1}$
is calculated in the energy band 2-30 keV. Errors are quoted at a
90\% confidence level.  The letters a, b, c and d stand for the
segments in the light curve.}
\end{table*}


\begin{thebibliography}{}

\bibitem[Altamirano et al.(2008)]{alt08} Altamirano, D., van der Klis, M., M{\'e}ndez, M., et al. 2008, \apj, 685, 436
\bibitem[Belloni(2010)]{bel10} Belloni, T.~M. 2010, Lecture Notes in Physics, Berlin Springer Verlag, 794, 53
\bibitem[Belloni et al.(2007)]{bel07} Belloni, T., et al. 2007, \mnras, 379, 247
\bibitem[B{\"o}ttcher \& Liang(1999)]{bot99} B{\"o}ttcher, M., \& Liang, E.~P. 1999, \apjl, 511, L37
\bibitem[Casares et al.(2006)]{cas06} Casares, J., et al. 2006, \mnras, 373, 1235
\bibitem[Choudhury \& Rao(2004)]{cho04} Choudhury, M., \& Rao, A.~R. 2004, \apjl, 616, L143
\bibitem[Choudhury et al.(2005)]{cho05} Choudhury, M., et al. 2005, \apj, 631, 1072
\bibitem[Christian \& Swank(1997)]{chr97} Christian, D.~J., \& Swank, J.~H. 1997, \apjs, 109, 177
\bibitem[Corbet et al.(1986)]{cor86} Corbet, R.~H.~D.,  et al. 1986, \mnras, 222, 15P
\bibitem[di Salvo et al.(2009)]{dis09} di Salvo, T., D'A{\'{\i}}, A., Iaria, R., et al 2009, \mnras, 398, 2022
\bibitem[di Salvo et al.(2006)]{dis06} di Salvo, T., et al. 2006, ChJAAS, 6, 010000
\bibitem[Done \& Gierli{\'n}ski(2006)]{don06} Done, C., \& Gierli{\'n}ski, M. 2006, \mnras, 367, 659
\bibitem[Done et al.(2007)]{don07} Done, C., et al. 2007, \aapr, 15, 1
\bibitem[Done \& Diaz Trigo(2010)]{don10} Done, C., \& Diaz Trigo, M. 2010, \mnras, 407, 2287
\bibitem[Esin et al.(1997)]{esi97} Esin, A.~A., et al. 1997, \apj, 489, 865
\bibitem[Ford et al.(1998)]{for98} Ford, E.~C.,  et al. 1998, \apjl, 508, L155
\bibitem[Ford et al.(2000)]{for00} Ford, E.~C.,  et al. 2000, \apj, 537, 368
\bibitem[Galloway et al.(2008)]{gal08} Galloway, D.~K., et al. 2008, \apjs, 179, 360
\bibitem[Hasinger(1987)]{has87} Hasinger, G. 1987, IAU Symp.~125: The Origin and Evolution of Neutron Stars, 125, 333
\bibitem[Hasinger (1990)]{has90} Hasinger, G., 1990, Reviews of Modern Astronomy, 3, 60
\bibitem[Hasinger \& van der Klis(1989)]{has89} Hasinger, G., \& van der Klis, M. 1989, \aap, 225, 79
\bibitem[Hasinger et al.(1990)]{hasetal90} Hasinger, G., et al. 1990, \aap, 235, 131
\bibitem[Homan et al.(2010)]{hom10} Homan, J.,  et al. 2010, \apj, 719, 201
\bibitem[Lei et al.(2008)]{lei08} Lei, Y.J.,  et al. 2008, \apj, 677, 461
\bibitem[Li et al.(2007)]{li07} Li, S.L., et al. 2007, \apj, 666, 368
\bibitem[Lin et al.(2009)]{lin09} Lin, D., et al. 2009, \apj, 696, 1257
\bibitem[Liu et al.(2007a)]{liu07a} Liu, Q.Z., et al. 2007a, \aap, 469, 807
\bibitem[Liu et al.(2007b)]{liu07b} Liu, B.F., et al. 2007b, \apj, 671, 695
%\bibitem[Mendez et al.(1997)]{men97} Mendez, M.,  et al. 1997, \apjl, 485, L37
\bibitem[Meyer et al.(2007)]{mey07} Meyer, F., Liu, B.F., \& Meyer-Hofmeister, E.\ 2007, \aap, 463, 1
\bibitem[Miller et al.(2006)]{mil06} Miller, J.~M.,  et al. 2006, \apj, 653, 525
\bibitem[Mueck et al.(2011)]{mue11} Mueck, B., et al. 2011, The X-ray Universe 2011, 255
\bibitem[Nowak et al.(1999)]{now99} Nowak, M.~A., et al. 1999, \apj, 517, 355
\bibitem[Ng et al.(2010)]{ng10} Ng, C., et al. 2010, \aap, 522, A96
\bibitem[O'Brien et al.(2004)]{obr04} O'Brien, K., et al. 2004, \mnras, 350, 587
\bibitem[Popham \& Sunyaev(2001)]{pop01} Popham, R., \& Sunyaev, R 2001, AIP Conf.~Proc.~599: X-ray Astronomy: Stellar Endpoints,
AGN, and the Diffuse X-ray Background, 599, 870
\bibitem[Qu et al.(2001)]{qu01} Qu, J.~L., Yu, W., \& Li, T.~P. 2001, \apj, 555, 7
\bibitem[Remillard(2005)]{rem05} Remillard, R.~A. 2005,
in AIP Conf. Proc. 797, Interacting Binaries: Accretion,
Evolution, and Outcomes, ed. L. Burderi et al. (Melville, NY: AIP),231
\bibitem[Remillard \& McClintock(2006)]{rem06} Remillard, R.~A., \& McClintock, J.~E. 2006, \araa, 44, 49
\bibitem[Rykoff et al.(2007)]{ryk07} Rykoff, E.S., et al. 2007, \apj, 666, 1129
\bibitem[Smale et al.(1986)]{sma86} Smale, A.~P., et al. 1986, \mnras, 223, 207
\bibitem[Seon et al.(1997)]{seo97} Seon, K.I., et al. 1997, \apj, 479, 398
\bibitem[Sriram et al.(2007)]{sri07} Sriram, K., et al. 2007, \apj, 661, 1055
\bibitem[Sriram et al.(2009)]{sri09} Sriram, K., et al. 2009, RAA, 9, 901
\bibitem[Sriram et al.(2010)]{sri10} Sriram, K., et al. 2010, \apj, 725, 1317
\bibitem[Sriram et al.(2012)]{sri12} Sriram, K., et al. 2012, \apjs, 200, 16
\bibitem[Stella(1990)]{ste90} Stella, L. 1990, \nat, 344, 747
%\bibitem[van der Klis(1995)]{van95} van der Klis, M.\ 1995, in X-ray Binaries, ed.W. H. G. Lewin, J. van Paradijs, \&
%E. P. J. van den Heuvel (Cambridge: Cambridge Univ. Press), 252
\bibitem[van der Klis(2000)]{van00} van der Klis, M. 2000, ARA\&A, 38, 717
\bibitem[van der Klis(2006)]{van06}
van der Klis, M. 2006, %A review of rapid X-ray variability in X-ray binaries
 in Compact stellar X-ray sources, W.H.G. Lewin \& M. van
der Klis (eds. ), Cambridge University Press, p. 39
\bibitem[van Straaten et al.(2005)]{van05} van Straaten, S.,
van der Klis, M., \& Wijnands, R. 2005, \apj, 619, 455
\bibitem[Weng \& Zhang(2011)]{wen11} Weng, S.S., \& Zhang, S.N.  2011, \apj, 739, 42
\bibitem[Wijnands et al.(1998)]{wij98} Wijnands, R.,  et al. 1998, \apjl, 495, L39
\bibitem[Yu \& Dolence(2007)]{yu07} Yu, W., \& Dolence, J. 2007, \apj, 667, 1043
\bibitem[Yu \& Yan(2009)]{yu09} Yu, W., \& Yan, Z. 2009, \apj, 701, 1940
\bibitem[Zhang(2007)]{zha07} Zhang, C.\ 2007, Advances in Space Research, 40, 1480








\end{thebibliography}
\end{document}